\documentclass[reprint,superscriptaddress,aps,prb,twocolumn,floatfix]{revtex4-2}

\usepackage{tabularx}  
\makeatletter
\def\switch@array{}
\makeatother
\usepackage{setspace}
\usepackage{amsmath}
\usepackage{bm}
\usepackage{upgreek} 
\usepackage{graphicx}
\usepackage[nearskip,margin = 0pt]{subfig}
\usepackage{verbatim}
\usepackage{amsfonts}
\usepackage{amssymb}
\usepackage{textcomp}
\usepackage{mathrsfs}
\usepackage{mathtools}
\usepackage{booktabs}
\usepackage{pifont}
\usepackage{xcolor}
\usepackage{url}
\usepackage{caption}

\usepackage{xcolor}
\usepackage[colorlinks,linkcolor=blue,anchorcolor=blue,citecolor=blue,urlcolor=black]{hyperref}
\usepackage{ragged2e}
\DeclareGraphicsExtensions{.pdf,.eps,.png,.jpg,.mps}
\begin{document}

\title{ A 75-mL intrinsically stable atom-filtered laser enabling deployable quantum devices}
\author{Zijie Liu$^{1,2,*}$, Zheng Xiao$^{1,*}$, Zhiyang Wang$^{1}$, Xiaolei Guan$^{1}$, Xiaomin Qin$^{1}$, Suyang Wei$^{1}$, Hongtian Song$^{2,5}$, Baoshuai Wang$^{2,5}$, Jia Zhang$^{1}$, Xiaopeng Xie$^{1}$, Tiantian Shi$^{3,4,\dagger}$, Anhong Dang$^{1,\dagger}$, and Jingbiao Chen$^{1,3,6}$ \\
\vspace{3pt}
$^1$School of Electronics, Peking University, Beijing, 100871, China\\
$^2$CSG Electric Power Research Institute, Guangzhou, 510663, China\\
$^3$National Key Laboratory of Advanced Micro and Nano Manufacture Technology, School of Integrated Circuits, Peking University, Beijing, 100871, China\\
$^4$Beijing Advanced Innovation Center for Integrated Circuits, Beijing 100871, China\\
$^5$Guangdong Provincial Key Laboratory of Intelligent Measurement and Advanced  Metering of Power Grid, Guangzhou, 510663, China\\
$^6$Hefei National Laboratory, Hefei, 230088, China\\
\vspace{3pt}
Corresponding authors: $^\dagger$tts@pku.edu.cn, $^\dagger$ahdang@pku.edu.cn.\\
$^{*}$These authors contributed equally to this work.}



\date{\today}

\maketitle
\noindent
\large\textbf{Abstract} \\
\normalsize\textbf{
Numerous quantum devices require lasers strictly locked to atomic transitions. Atom-filtered lasers (AFLs) are considered a leading candidate for quantum device laser sources due to their ability to self-align to atomic transitions. However, the sub-GHz sharp transmission spectra of conventional atomic filters impose constraints on both laser miniaturization and output stability. Herein, we demonstrate a micro Faraday anomalous dispersion optical filter ($\bm{\upmu}$FADOF) operating within the extreme hyperfine Paschen-Back regime, generating a 7.5 GHz flat-top transmission window. By integrating this filter, the miniaturization bottleneck of the AFLs is overcome, achieving a 30-fold volume reduction to a compact package volume of 75 mL. Simultaneously, investigations into the optical self-feedback characteristics of the $\bm{\upmu}$FADOF reveal a stable operating regime, where the optical self-feedback of the atomic filter acts as a stabilizing mechanism, enabling the laser to achieve a power instability of $\mathbf{9 \times 10^{-6}}$ at 1 s and $\mathbf{3.2 \times 10^{-5}}$ at 8400 s. The optical frequency standard based on this 75-mL AFL achieves a further improvement in long-term frequency stability to $\mathbf{3\times10^{-13}}$ at 10000 s and maintains turn-key operation under environmental shocks, highlighting the critical role of this laser in the development of deployable quantum devices.}

\vspace{3pt}
\maketitle
\noindent
\large\textbf{Introduction} \\
\normalsize
\noindent 
\begin{figure*}[ht]
\centering
\captionsetup{singlelinecheck=no, justification = RaggedRight}
\includegraphics[width=17cm]{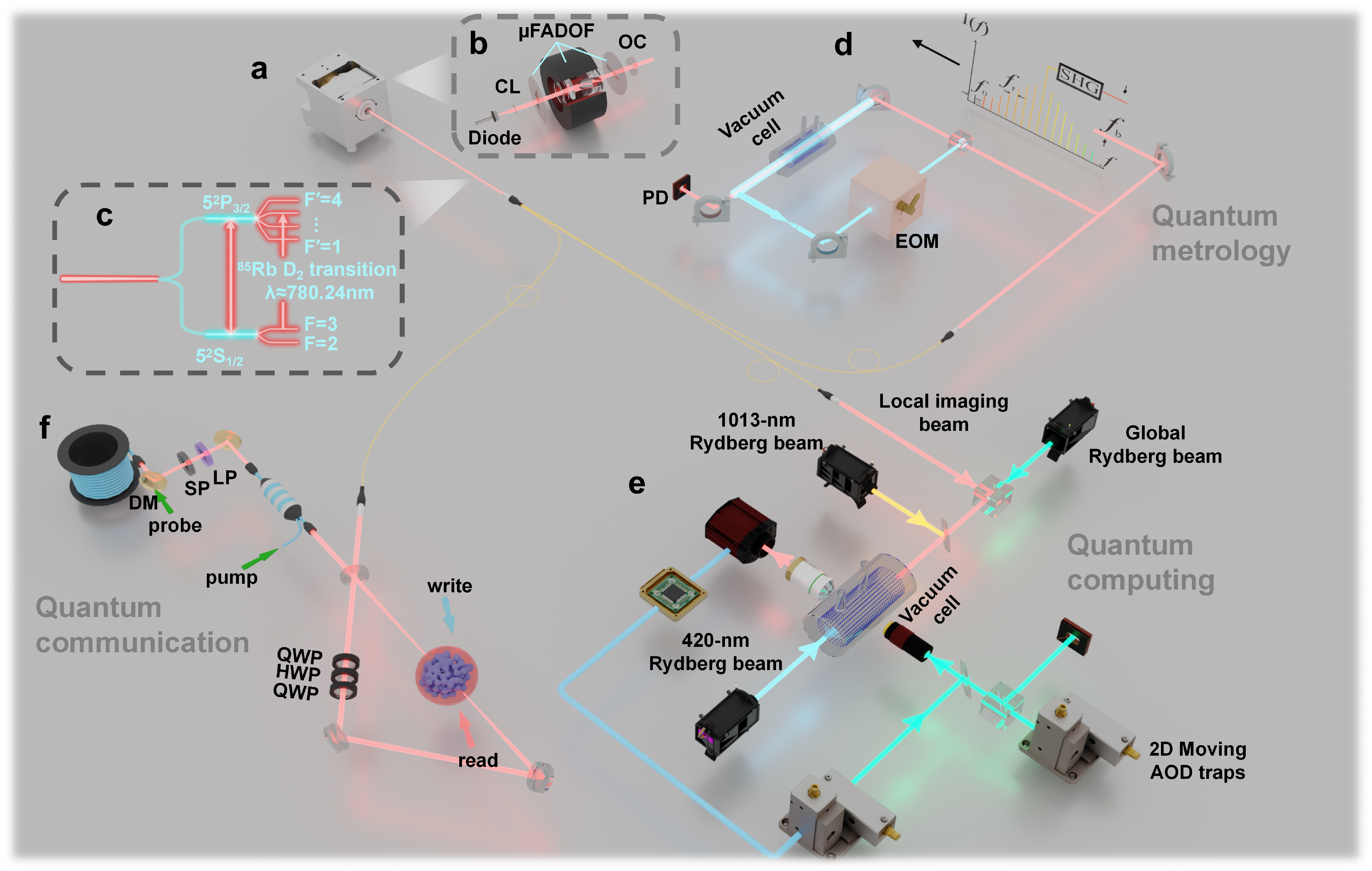}
\caption{\textbf{75-mL atom-filtered laser (75-mL AFL) and its applications in deployable quantum devices. a}. External structure of the 75-mL AFL. \textbf{b}. Internal structure of the 75-mL AFL, in which a micro $^{85}$Rb Faraday anomalous dispersion optical filter ($\bm{\upmu}$FADOF) is integrated inside the laser resonator. CL, collimating lens; OC, output coupler. \textbf{c}. The 75-mL AFL output laser is directly tuned to the $^{85}$Rb $5^2$S$_{1/2} \rightarrow 5^2$P$_{3/2}$ transition. \textbf{d}. Quantum metrology application: a compact optical atomic clock based on modulation transfer spectroscopy (MTS) frequency locking, in which the 75-mL AFL serves as the local oscillator laser. A erbium-doped optical frequency comb, which has undergone frequency doubling through a second harmonic generation (SHG) crystal, converts the optical frequency instability into microwave-frequency instability. PD, photondiode \textbf{e}. Quantum computing application: the 780-nm output of the 75-mL AFL serves as the local imaging beam of a neutral-atom quantum computer to realize a programmable quantum processor based on encoded logical qubits. \textbf{f}. Quantum communication application: at a memory node in long-distance matter-matter entanglement, the 75-mL AFL output laser can serve as the write and read beams for spin-wave writing and readout. DM, dichroic mirror; LP(SP), long(short)-pass filter; H(Q)WP, half(quarter)-wave plate.}
\label{fig1}
\end{figure*}
\begin{figure*}[ht]
\centering
\captionsetup{singlelinecheck=no, justification = RaggedRight}
\includegraphics[width=17cm]{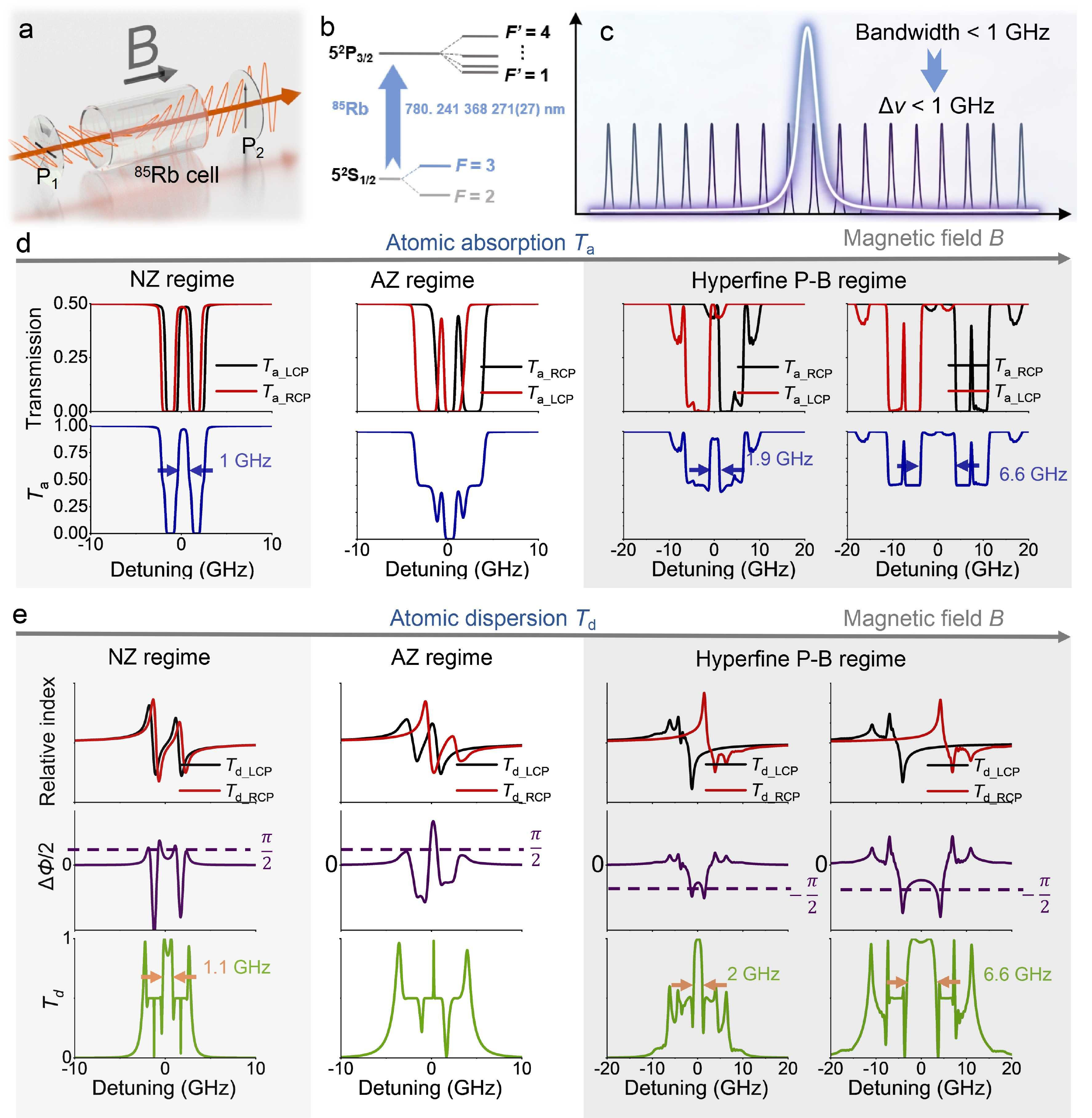}
\caption{\textbf{Working principle of FADOF frequency selection and flat-top FADOF design. a,} Schematic diagram illustrating the frequency selection principle of the FADOF. \textbf{b,} Energy-level diagram of the $^{85}$Rb $5^2$S$_{1/2} \rightarrow 5^2$P$_{3/2}$ transition. \textbf{c,} Frequency selection principle of the AFL. Only the laser resonator modes (blue lines) located within the FADOF transmission spectrum (white line) can oscillate. \textbf{d,} The left-hand circularly polarized (LCP, red curves) and right-hand circularly polarized (RCP, black curves) components of the atomic absorption term $T_a$ (blue curves) across different magnetic field regimes. \textbf{e,} The LCP (black curves) and RCP (red curves) components of the atomic dispersion term $T_d$ (green curves) across different magnetic field regimes. Subtracting the two dispersion curves yields the optical rotation angle profile (purple curves) of the laser after traversing the atomic filter, which is then used to derive the atomic dispersion term profile (green curves). The magnetic field strengths are 100 G in the normal Zeeman regime (NZ regime), 500 G in the anomalous Zeeman regime (AZ regime), and 2500 G and 3500 G (from left to right) in the hyperfine Paschen-Back regime (hyperfine P-B regime).} 
\label{fig2}
\end{figure*}
In recent years, quantum technology has undergone rapid development, demonstrating transformative significance across diverse fields—such as computing \cite{bernien2017probing,bluvstein2024logical,ebadi2021quantum}, communication \cite{luo2026entangling,ritter2012elementary}, and precision metrology \cite{sedlacek2012microwave,roslund2024optical}—and continuously catalyzing breakthrough discoveries \cite{bothwell2022resolving,collaboration2021frequency,pedrozo2020entanglement}. In quantum computing, logical quantum processors based on reconfigurable neutral-atom arrays have made remarkable strides, successfully realizing systems containing up to 280 physical qubits \cite{bluvstein2024logical}. In the field of quantum communication, researchers have utilized laser-cooled atomic ensembles as quantum memories to achieve optical-fiber entanglement distribution over a distance of 420 kilometers \cite{luo2026entangling}. In the realm of quantum metrology, a novel method for microwave electrometry has been realized using rubidium ($^{87}$Rb) Rydberg atoms excited within a glass vapor cell, achieving a remarkable measurement sensitivity of $\sim30\text{ }\mu\text{V cm}^{-1}\text{Hz}^{-1/2}$ \cite{sedlacek2012microwave}. Across all these frontier achievements, atomic transition lines provide stable and accurate "reference scales" in nature. However, the use of atomic transitions imposes stringent requirements: the frequency of the laser source must remain strictly locked to specific atomic transition lines.

Traditional solutions employ grating- or etalon-based external-cavity diode lasers (ECDLs) as seed sources \cite{r4,zorabedian1988interference,r6,hilton2025demonstration,takamoto2020test,zhang2022device}, where the laser frequency is aligned to the atomic transition via coarse mechanical tuning of intracavity frequency-selective elements, in conjunction with fine tuning of the diode current and temperature. Under long-term environmental disturbances, the frequency of these lasers tends to drift away from the atomic transition, leading to quantum device failure or operational errors \cite{hilton2025demonstration}. This precise yet fragile nature has long confined quantum devices to tightly controlled laboratory environments, hindering their deployment in demanding field conditions.

The atom-filtered lasers (AFLs) \cite{r18,r27,keaveney2016single,chang2022frequency,qin2024switchable,liu2025turn,wei2026dual} have addressed this issue, by integrating a Faraday anomalous dispersion optical filter (FADOF) \cite{ohman1956some,menders1991ultranarrow,portalupi2016simultaneous} into the laser resonator. The laser output mode is confined within the transmission bandwidth of the FADOF, enabling the laser frequency to remain continuously aligned with the atomic transition line, immune to long-term external perturbations. While AFLs have shown great promise in atomic devices, their deployment in complex field environments (e.g., satellite or vehicular platforms) is severely hindered by their bulky size and unstable laser output \cite{liu2025turn}. The root of these challenges lies in the filter's sub-GHz transmission bandwidth \cite{keaveney2016single,chang2022frequency}, its steeply varying Lorentzian-like profile \cite{r18,r27}, and optical self-feedback \cite{qin2024switchable,liu2025turn}. Specifically, ensuring that at least one longitudinal mode lies within the effective gain window requires the longitudinal mode spacing to be no greater than the window width. Consequently, the \textless 1 GHz transmission window restricts the laser cavity length to \textgreater 15 cm, fundamentally precluding device miniaturization. Furthermore, the \textless 1 GHz bandwidth, coupled with the intensity-sensitive optical self-feedback, renders the laser susceptible to frequent mode-hopping and severe output power instability when placed in complex environments.

Herein, we demonstrate an intensity-insensitive, micro Faraday anomalous dispersion optical filter ($\upmu$FADOF) featuring an ultra-wide 7.5 GHz flat-top transmission profile, pushing the atomic filter out of the conventional normal and anomalous Zeeman regimes (typically below hundred Gauss) \cite{zeeman1897xxxii,bohr1926spinning}and into the extreme hyperfine Paschen-Back regime \cite{back1928kernmoment,sargsyan2014hyperfine}. The transmission window bandwidth is broadened from less than 1 GHz to 7.5 GHz, enabling a minimum laser cavity length of 2 cm, thereby breaking through the physical limits of AFL miniaturization and reducing the total volume by a factor of 30 to a mere $4.65 \times 3.9 \times 4.15\text{ cm}^3$. Furthermore, based on an in-depth investigation into the optical self-feedback mechanism of the $\upmu$FADOF, we categorize its operating states into the saturable absorption, stable, and reverse saturable absorption regimes according to the operating optical intensity. By setting the AFL to work in the stable regime, the self-feedback of $\upmu$FADOF suppresses rather than exacerbates the output power fluctuations, thereby improving the power instability by two to three orders of magnitude to achieve $9 \times 10^{-6}$ at 1 s and $3.2 \times 10^{-5}$ at 8400 s.

To fully demonstrate the potential applications of the 75-mL AFL, we deployed it as the local oscillator in a compact rubidium optical frequency standard. Benefiting from the long-term power stability of the 75-mL AFL, the optical frequency standard achieved a further improvement over the best previously reported long-term stability for Rb MTS frequency locking, reaching a record $3 \times 10^{-13}$ at 10000 s. Most importantly, in rigorous environmental and vibrational shock experiments, this frequency standard exhibited excellent "turn-key" characteristics, demonstrating the significant potential of the 75-mL AFL in the development of out-of-lab quantum devices. Fig. \ref{fig1} illustrates the near-term prospective application domains for the 75-mL AFL.

\vspace{6pt}
\noindent
\large\textbf{Results}\\
\noindent
\normalsize
\textbf{7.5 GHz flat-top transmission window $\bm{\upmu}$FADOF}\\
\noindent 
Fig. \ref{fig2}a illustrates the schematic of the Faraday anomalous dispersion optical filter (FADOF). Upon passing through the polarizer P$_1$, the incident light is converted into linearly polarized light with a degree of polarization exceeding 99\%. As this beam propagates through an atomic vapor cell subjected to a longitudinal magnetic field, the dispersion curves for its left-hand circularly polarized (LCP) and right-hand circularly polarized (RCP) components split due to the Zeeman effect, inducing a refractive index difference $\Delta n$. While traversing the cell length $L$, this refractive index difference accumulates into a phase difference $\Delta \phi=2\pi \Delta n L/\lambda_0$, causing the polarization plane of the recombined linearly polarized light to rotate relative to the incident beam, where $\lambda_0$ is the laser wavelength. The closer this rotation angle $\Delta \phi/2$ approaches $\pi/2$, the higher the laser's transmittance through the analyzer P$_2$ (which is oriented orthogonally to the polarizer P$_1$). As the laser frequency approaches the atomic transition, the Zeeman-induced dispersion splitting becomes more pronounced, making it easier for the rotation angle to reach $\pi$/2. Consequently, the FADOF's transmission windows are localized around atomic transition lines; for an $^{85}$Rb FADOF, its sub-GHz-bandwidth transmission window is typically centered near 780.241 nm, as depicted in Fig. \ref{fig2}b.

Integrating the FADOF into the resonator of a semiconductor laser suppresses all laser modes outside the transmission window, ensuring laser frequency self-alignment to atomic transition lines. Although such AFLs have been widely adopted in quantum device development due to their immunity to long-term drift and inherent alignment to atomic transitions, the inability to be miniaturized and unstable laser emission remain the primary bottlenecks preventing AFL-based quantum devices from moving out of the laboratory. As depicted in Fig. \ref{fig2}c, the FADOF's sub-GHz sharp transmission window is the critical factor limiting miniaturization. To sustain normal AFL operation, at least one longitudinal laser mode must fall within this transmission window. This imposes a requirement on the longitudinal mode spacing of $\Delta \nu$ \textless 1 GHz, which dictates that the laser cavity length must be $L = c / 2\Delta\nu$ \textgreater 15 cm, fundamentally restricting laser miniaturization. Furthermore, as the laser mode drifts within the sharp window, the FADOF's transmittance fluctuates significantly, degrading the laser's output power stability. Therefore, developing a FADOF with an ultrabroadband, flat-top transmission window is a prerequisite for the large-scale practical deployment of AFLs.

\begin{table*}[tbp]
\captionsetup{singlelinecheck=no}
\caption{Required magnetic field and transmission bandwidth of different atomic species} 
\label{tab:atomic_params}
\begin{tabularx}{\textwidth}{c >{\centering\arraybackslash}X >{\centering\arraybackslash}X >{\centering\arraybackslash}X}
\toprule
Atomic Species & Required Magnetic Field & Predicted Transmission Bandwidth & Measured Transmission Bandwidth \\
\midrule
$^{85}$Rb & $>3500$ G & $>6.6$ GHz & $7.5$ GHz \\
$^{87}$Rb & $>5000$ G & $>7$ GHz & To be measured \\
Cs & $>7000$ G & $>11$ GHz & $13$ GHz \\
\bottomrule
\end{tabularx}
\end{table*}

There have been few studies detailing the expansion of atomic filter transmission bandwidths. While the prevalent method is the addition of buffer gases at varying pressures, its broadening effect is limited and comes at the cost of severely degraded transmittance, rendering it unviable for AFL optimization. We attempt to break beyond the traditional magnetic field limits of atomic filters (i.e., the normal and anomalous Zeeman regimes) to operate within the hyperfine Paschen-Back regime. A strong magnetic field is utilized to thoroughly pull apart the RCP and LCP components of the absorption and dispersion profiles in the frequency domain, thereby generating a large-bandwidth, flat-top transmission window at the center frequency. The underlying principle is depicted in Figs. \ref{fig2}d and \ref{fig2}e. Building upon the theory of dispersive magneto-optical filters \cite{yeh1982dispersive}, we establish that the overall atomic filter transmittance $T$ is governed by both atomic absorption and dispersion, expressed as:
\begin{equation}
T = T_a \times T_d
\label{eq:1} 
\end{equation}
where $T_a$ and $T_d$ denote the respective contributions from atomic absorption and atomic dispersion. The detailed theoretical analysis is presented in the \textbf{Theoretical calculations} of the \textbf{Methods}. In addition, theoretical models accounting for the Doppler effect and the multi-level structure are also considered \cite{xiao2026statedependentdiffusionspectrastrongly}.

Regarding the atomic absorption term, as shown in Fig. \ref{fig2}d, its RCP and LCP components split under the influence of the magnetic field. When the magnetic field operates in the normal or anomalous Zeeman regime, the splitting between the two components is relatively small and their profiles overlap. Consequently, the combined atomic absorption term lacks a broad, near-zero-absorption transmission window around the atomic transition. However, as the magnetic field breaks into the Paschen-Back regime, the RCP and LCP components become completely separated in the frequency domain, leading to the emergence of a transmission window at the line center. According to the hyperfine Paschen-Back effect, increasing the magnetic field causes the different magnetic sublevels of the atom to shift further apart. This results in a nearly linear frequency-domain separation between the RCP and LCP components, which in turn approximately linearly broadens the central window. With a sufficiently strong magnetic field, a transmission window with an adequately large bandwidth can be generated at the atomic transition.

As for the atomic dispersion term (Fig. \ref{fig2}e), the RCP and LCP components similarly split under the magnetic field. The phase difference profile can be obtained via the formula $\Delta \phi = 2\pi \Delta n L / \lambda_0$, and the atomic dispersion term $T_d$ is proportional to a sine function of this phase difference. In the normal or anomalous Zeeman regime, the small component splitting prevents $T_d$ from forming a broad transmission window. However, upon entering the hyperfine Paschen-Back regime, the slopes of the two dispersion curves become comparable at the transition center. The resulting phase difference profile obtained from their subtraction exhibits a slope approaching zero, thereby realizing a flat-top $T_d$ transmission profile. Similar to the atomic absorption term, a stronger magnetic field yields a broader transmission window. For the $^{85}$Rb atom at a magnetic field of 3550 G, combining Figs. 2d and 2e with Equation \ref{eq:1} indicates that the atomic filter possesses a flat-top transmission window with a bandwidth of 6.6 GHz.

The aforementioned FADOF design principle is universally applicable to other alkali metal atoms, including Cs, $^{85}$Rb, and $^{87}$Rb. We predicted and experimentally verified the magnetic field requirements and achievable bandwidths for practical flat-top FADOFs based on these different atoms, as summarized in Table \ref{tab:atomic_params}. Because $^{85}$Rb requires the lowest magnetic field—thereby facilitating device miniaturization—it was selected for the construction of the 75-mL AFL in this study.

\vspace{3pt}
\noindent 
\maketitle
\textbf{Optical self-feedback of $\bm{\upmu}$FADOF}\\
\noindent 
\begin{figure*}[ht]
\centering
\captionsetup{singlelinecheck=no, justification = RaggedRight}
\includegraphics[width=17cm]{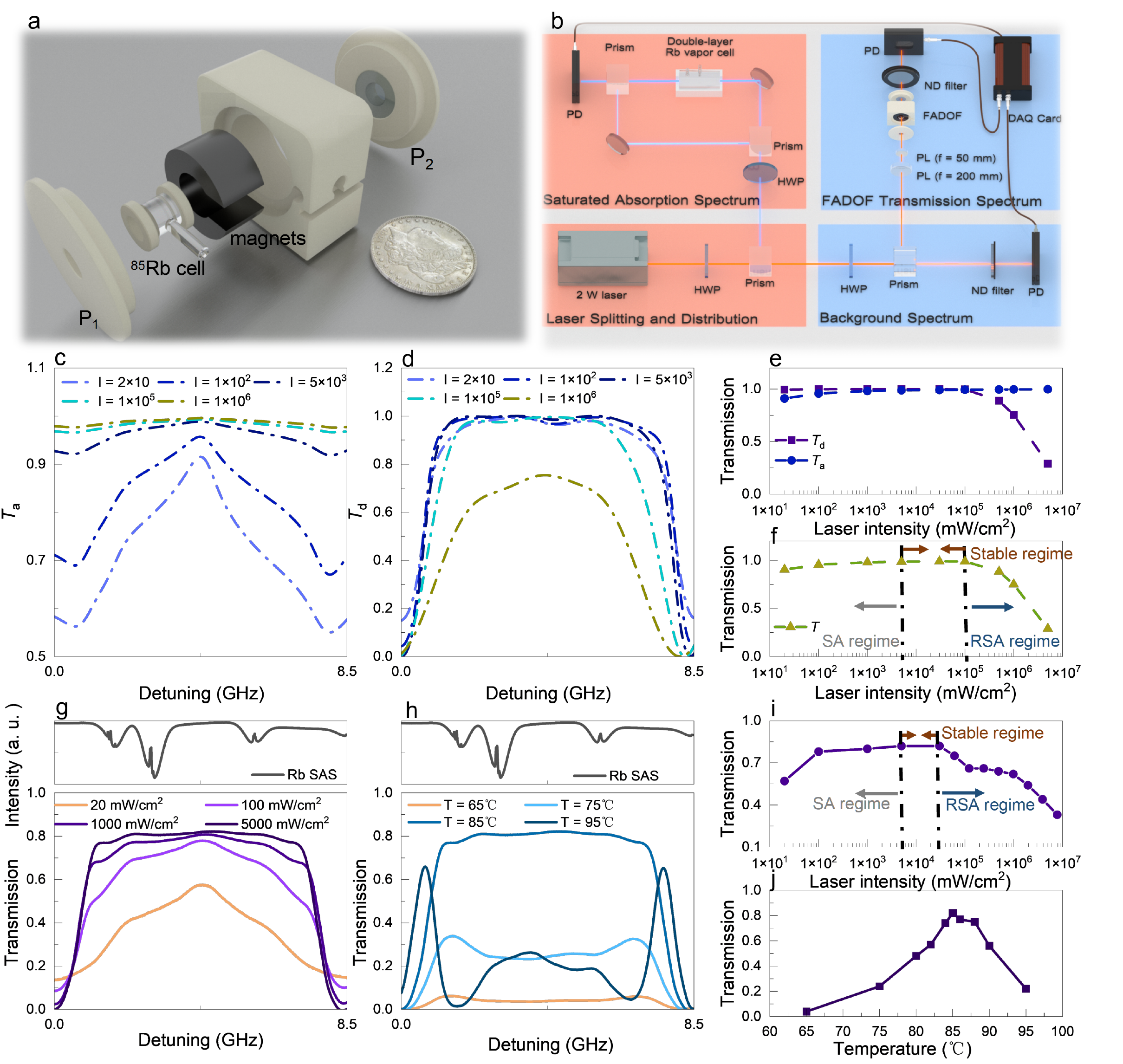}
\caption{\textbf{$\upmu$FADOF device, test system, optical feedback characteristics, and temperature characteristics. a,} Schematic of the $\upmu$FADOF device. The cold finger of the atomic vapor cell, the two probes of the temperature control system, and the power supply lines are all positioned within the gap of the C-shaped magnet. Two polarizers, each 0.1 mm thick, are placed in front of and behind the atomic vapor cell, functioning as the polarizer and analyzer for the FADOF. \textbf{b,} The $\upmu$FADOF transmission spectrum measurement system, which is divided into four functional sections. \textit{Laser splitting and distribution}: A small portion (a few mW) of the test laser is utilized for saturated absorption spectroscopy, while the majority is directed toward the FADOF transmission spectrum measurement. \textit{Saturated absorption spectroscopy}: The natural rubidium saturated absorption spectrum is scanned simultaneously with the transmission spectrum to serve as a frequency reference. \textit{Background spectrum}: The laser power spectrum is recorded prior to entering the FADOF to eliminate inherent power fluctuations caused by frequency scanning. \textit{FADOF transmission spectrum}: The bulk of the test laser (up to 1.8 W) is beam-compressed to a cross-section of $0.14 \times 0.21$ mm$^2$ before entering the $\upmu$FADOF to acquire transmission spectrum data, reaching a maximum test optical intensity of $7.8 \times 10^6$ mW/cm$^2$. \textbf{c,d,e,f,} Theoretical spectrum for the atomic absorption component $T_a$ (\textbf{c}) and the atomic dispersion component $T_d$ (\textbf{d}) of the $\upmu$FADOF transmission spectrum under varying optical intensities. By analyzing the relationship between the center transmittance and the optical intensity, their respective optical feedback characteristics are derived (\textbf{e}), which subsequently yields the overall optical feedback characteristics of the $\upmu$FADOF (\textbf{f}) under the combined effect of both components. \textbf{g,} Experimental transmission spectrum of the $\upmu$FADOF under different optical intensities within the saturable absorption (SA) regime (0 to $5 \times 10^3$ mW/cm$^2$). \textbf{h,} Experimental transmission spectrum of the $\upmu$FADOF at different temperatures under a fixed optical intensity of $5 \times 10^3$ mW/cm$^2$. \textbf{i,} Experimental results characterizing the optical self-feedback of the $\upmu$FADOF. \textbf{j,} Experimental results of the $\upmu$FADOF temperature characteristics.}

\label{fig3}
\end{figure*}
\begin{figure*}[ht]
\centering
\captionsetup{singlelinecheck=no, justification = RaggedRight}
\includegraphics[width=17cm]{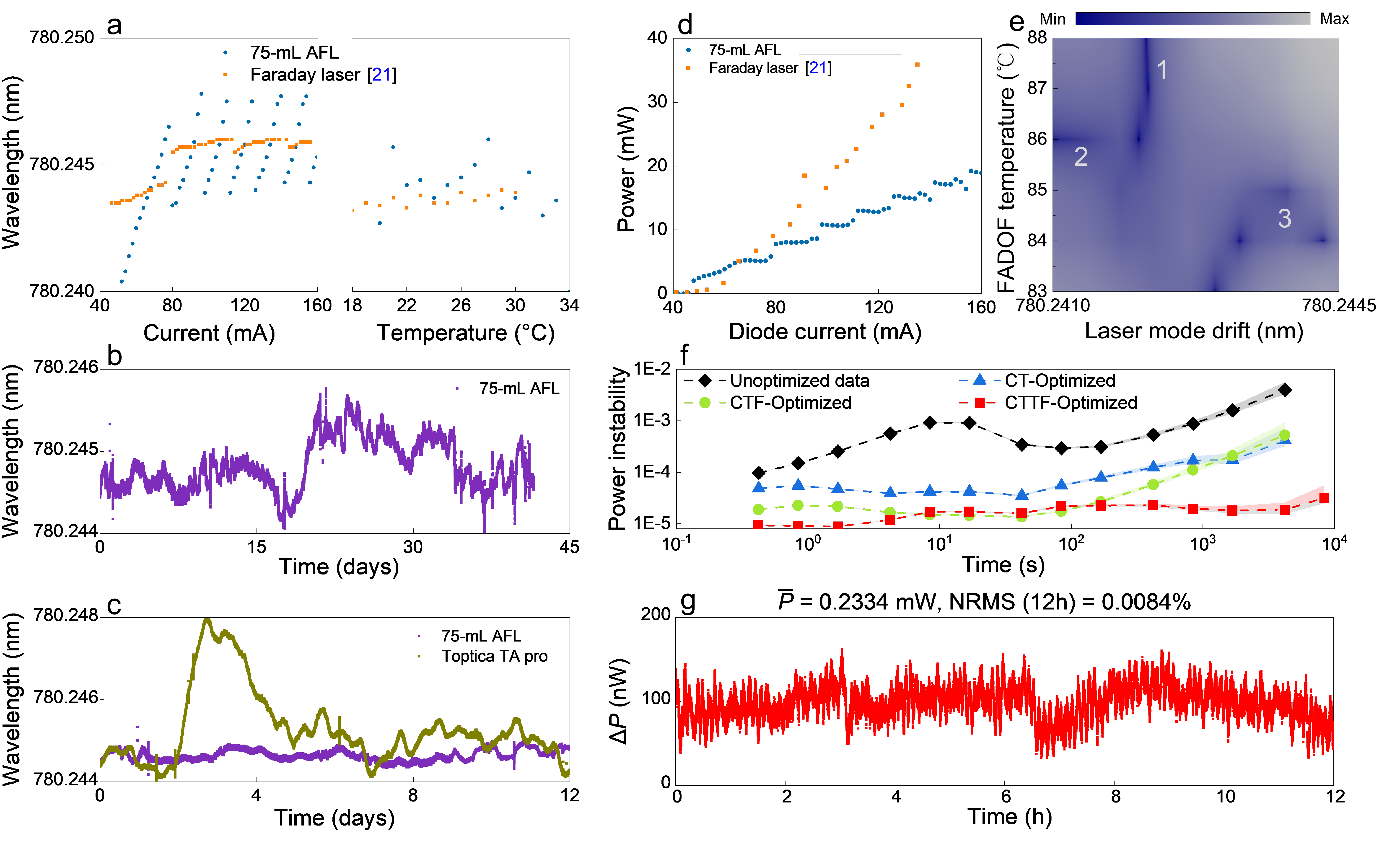}
\caption{\textbf{Wavelength and power characteristics of the 75-mL AFL. a,} Output wavelengths of the 75-mL AFL and previous Faraday lasers \cite{qin2024switchable} under varying diode currents and temperatures. \textbf{b,} Wavelength drift of the 75-mL AFL over a 42-day free-running period. \textbf{c,} Wavelength drift of the 75-mL AFL compared with an established commercial laser (Toptica TA pro) operating simultaneously in free-running mode for 12 days in the same laboratory environment. \textbf{d,} Output power of the 75-mL AFL and previous Faraday lasers \cite{qin2024switchable} under varying diode currents. \textbf{e,} The rate of power change at different FADOF temperatures and laser wavelengths during the power optimization process (darker colors indicate lower rates of change). \textbf{f,} Power instability (Allan deviation) achieved by adopting various optimization measures. C: optimization of the diode current parameters; First T: optimization of the FADOF temperature parameters; Second T: overall active temperature control and optimal parameter selection for the 75-mL AFL package; F: laser frequency locking. \textbf{g,} Power fluctuations of the 75-mL AFL during a 12-hour free-running period. The sampled optical power is 0.2334 mW; NRMS denotes the normalized root-mean-square of laser power.}
\label{fig4}
\end{figure*} 
\begin{figure*}[ht]
\centering
\captionsetup{singlelinecheck=no, justification = RaggedRight}
\includegraphics[width=17cm]{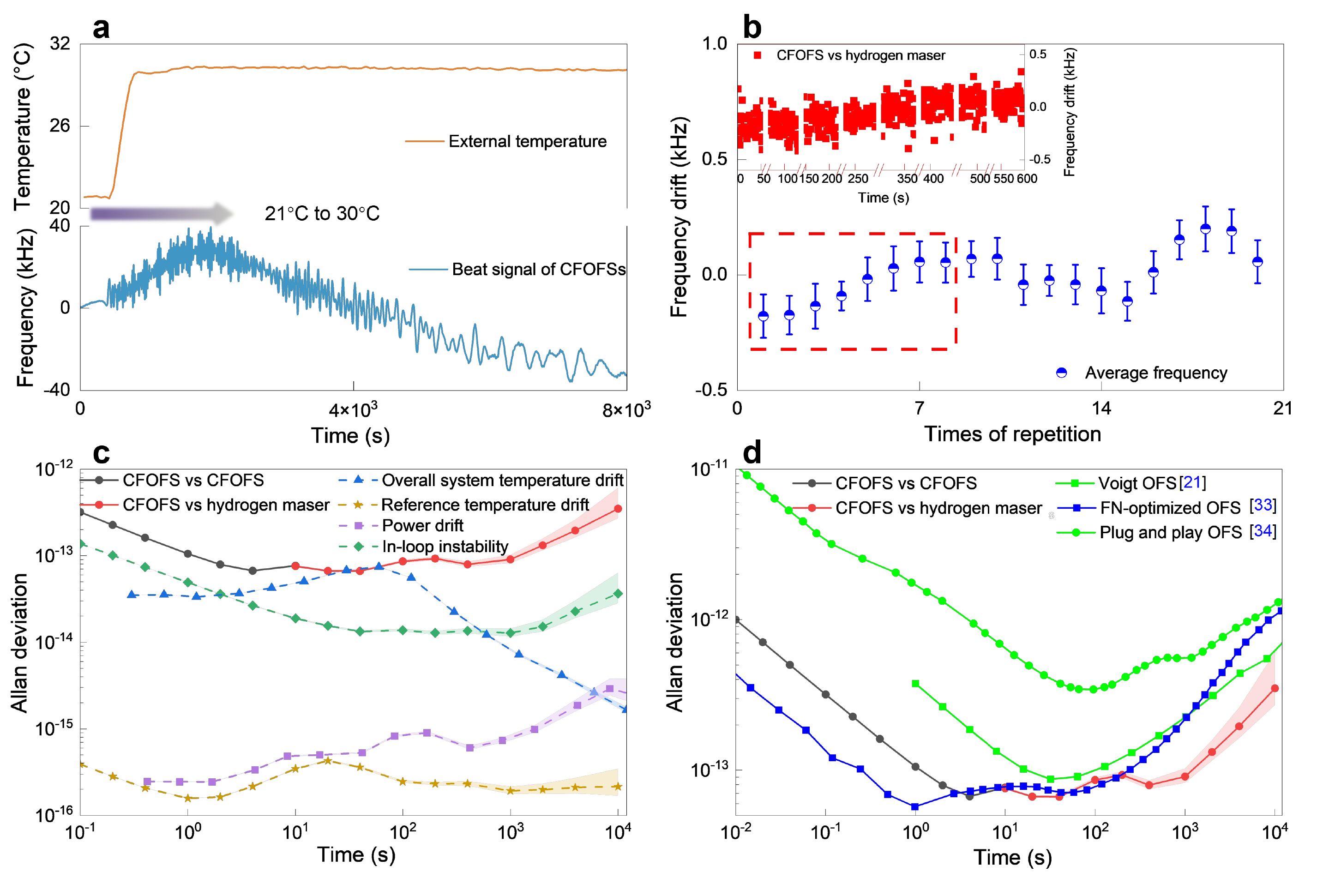}
\caption{\textbf{Turn-key characteristics and frequency instability of the CFOFS. a,} Frequency drift of the beat signal between two CFOFSs when subjected to an ambient temperature shock of +9°C within 6 minutes. \textbf{b,} Laser frequency drift during 20 vibration shock tests. The inset displays the CFOFS frequency output during the first 8 vibration shocks; after each shock, the data is averaged over 1 minute and recorded before the subsequent shock is applied. \textbf{c,} CFOFS frequency instability and the noise contributions from individual parameters. \textit{CFOFS vs CFOFS}: Short-term frequency instability evaluated via the beat note between two CFOFSs. \textit{CFOFS vs hydrogen maser}: long-term frequency instability evaluated via the beat note between a CFOFS and an optical frequency comb locked to a hydrogen maser. \textit{Overall system temperature drift}: Frequency noise induced by the thermal drift of the entire optical module. \textit{Reference temperature drift}: Frequency noise induced by the temperature drift of the reference cell. \textit{Power drift}: Frequency noise induced by the power drift of the 75-mL AFL. \textit{In-loop instability}: Internal noise of the phase-locked loop (PLL). \textbf{d,} Comparison of the CFOFS frequency instability with previously reported compact MTS optical frequency standards (Voigt OFS \cite{liu2025turn}, FN-optimized OFS \cite{lee2023laser} and Plug and play OFS \cite{strangfeld2022compact}).}
\label{fig5}
\end{figure*}
A 7.5 GHz flat-top transmission spectrum is provided by a $^{85}$Rb $\upmu$FADOF operating in the hyperfine Paschen-Back regime, with its configuration shown in Fig. 3a. A $^{85}$Rb atomic vapor cell (length: 10 mm, diameter: 7 mm, purity: 99.9\%) wrapped in a heating film is mounted inside a length-matched C-shaped magnet, which provides an inhomogeneous magnetic field (1000–3550 G)along the length of the vapor cell. Affected by the internal magnetic field inhomogeneity, the transmission window of the $\upmu$FADOF deviates significantly from theoretical expectations, as indicated by the orange line in Fig. \ref{fig3}g. To resolve this issue, we incorporated the influence of optical intensity on the transmission window and investigated the optical self-feedback mechanism of the atomic filter.

We analyze the atomic absorption term $T_a$ and the atomic dispersion term $T_d$ of the $\upmu$FADOF transmission spectra under a non-uniform magnetic field at varying optical intensities, as shown in Figs. \ref{fig3}c and \ref{fig3}d. At a low optical intensity (20 mW/cm$^2$), the non-uniform magnetic field distorts the atomic absorption term $T_a$ from the 6.6 GHz flat-top transmission window (seen in Fig. \ref{fig2}d) into a 3 GHz triangular-shaped transmission window, as indicated by the light blue curve in Fig. \ref{fig3}c. As the optical intensity increases, the atomic absorption gradually saturates. As depicted in Fig. \ref{fig3}c, $T_a$ eventually becomes transparent across the entire frequency domain and ceases to influence the final transmission window of the $\upmu$FADOF. The transmittance of $T_a$ at the center frequency of the atomic transition as a function of optical intensity is plotted as the blue curve in Fig. \ref{fig3}e. The analysis indicates that for optical intensities exceeding $5 \times 10^3$ mW/cm$^2$, $T_a$ reaches saturation and remains essentially constant.

Meanwhile, the dependence of the atomic dispersion term $T_d$ on optical intensity is illustrated in Fig. \ref{fig3}d. For intensities below $1 \times 10^5$ mW/cm$^2$, the spectral profile remains largely unaffected; however, once the intensity surpasses $1 \times 10^5$ mW/cm$^2$, $T_d$ is suppressed, resulting in a decrease in both transmittance and transmission bandwidth. The transmittance of $T_d$ at the atomic transition center frequency is shown by the purple curve in Fig. \ref{fig3}e, confirming that $T_d$ begins to decline at intensities greater than $1 \times 10^5$ mW/cm$^2$.

Taking both the atomic absorption term $T_a$ and the atomic dispersion term $T_d$ into account, we propose that the optical self-feedback characteristics of the $\upmu$FADOF can be classified into three distinct regimes, as illustrated in Fig. \ref{fig3}f. At optical intensities below $5 \times 10^3$ mW/cm$^2$, the variation in $T_a$ dominates. The $\upmu$FADOF transmittance increases with optical intensity, exhibiting saturable absorption characteristics; this is termed the saturable absorption regime (SA regime). For optical intensities between $5 \times 10^3$ mW/cm$^2$ and $1 \times 10^5$ mW/cm$^2$, both $T_a$ and $T_d$ remain constant, and the $\upmu$FADOF transmission spectrum stably maintains a 7.5 GHz flat-top profile; this is defined as the stable regime. Finally, when the optical intensity exceeds $1 \times 10^5$ mW/cm$^2$, the contribution of $T_d$ becomes dominant. In this regime, the $\upmu$FADOF transmittance decreases as the optical intensity rises, displaying reverse saturable absorption characteristics; this is designated as the reverse saturable absorption regime(RSA regime). A detailed theoretical derivation of the optical self-feedback mechanism of the $\upmu$FADOF is presented in the \textbf{Theoretical calculations} of the \textbf{Methods}.

A transmission spectrum measurement system for the $\mu$FADOF was constructed, as illustrated in Fig. \ref{fig3}b. This setup utilizes the saturated absorption spectrum of Rb atoms as the frequency reference for the $\mu$FADOF and employs a tapered amplifier (TA) boosted grating-based ECDL with a 2 W output power as the test light source. The tested optical intensity ranges from 2 to $7.8 \times 10^6$ mW/cm$^2$. The experimental results verifying the optical self-feedback characteristics of the $\upmu$FADOF are shown in Figs. \ref{fig3}g and \ref{fig3}i. Specifically, when the $\upmu$FADOF operates in the SA regime, the evolution of its transmission spectrum (Fig. \ref{fig3}g) demonstrates that the transmission window distortion induced by the magnetic field inhomogeneity is gradually eliminated as the optical intensity increases. Once the optical intensity reaches $5 \times 10^3$ mW/cm$^2$, the $\upmu$FADOF enters the stable regime. Here, the transmission window recovers its 7.5 GHz flat-top profile and remains constant across a subsequent range of optical intensities. The corresponding transmittance at the center frequency of the atomic transition as a function of optical intensity is plotted in Fig. \ref{fig3}i. The observed trend closely mirrors the theoretical prediction (Fig. \ref{fig3}f), identically dividing into the SA, stable, and RSA regimes. The primary discrepancy is that the experimentally measured upper intensity limit for the stable regime is $3 \times 10^4$ mW/cm$^2$, which is lower than the theoretically derived $1 \times 10^5$ mW/cm$^2$. This deviation is primarily attributed to the axial inhomogeneity of the optical intensity—caused by the divergence of the focused laser beam within the atomic vapor cell during the measurement—as well as discrepancies between the actual and simulated magnetic fields.

Furthermore, investigating the temperature characteristics of the $\mu$FADOF operating in the stable regime reveals a correlation between the optimal operating temperature and the applied optical intensity. Briefly, for every twofold increase in optical intensity, the operating temperature of the $\mu$FADOF must be raised by 2°C to maintain an unaltered transmission window profile and transmittance. At an optical intensity of $5 \times 10^3$ mW/cm$^2$, the dependencies of the $\mu$FADOF spectral profile and maximum transmittance on temperature are shown in Figs. \ref{fig3}h and \ref{fig3}j, respectively, demonstrating that both the flat-top profile and peak transmittance are achieved at 85°C.

Based on the above analysis, the optimal operating parameters for the $\upmu$FADOF are 85 °C and 5000 mW/cm$^{2}$. Under these conditions, the 75-mL AFL achieves high output power and optimal power stability.

\vspace{3pt}
\noindent
\textbf{75-mL Atom-filtered laser}
\noindent\\
As shown in Fig. 1, the 75-mL AFL integrates a diode chip, $\upmu$FADOF, their independent temperature control modules, laser cavity mirrors, and other components onto an integrated invar base—forming a straight-cavity AFL with a 3.3 cm cavity length and $4.65\times3.9\times4.15$ cm$^{3}$ volume (1/30 that of conventional AFLs). The 75-mL AFL inherits AFL characteristics \cite{chang2022frequency,qin2024switchable,liu2025turn}: (1) The laser frequency self-aligns to the atomic transition lines immediately upon startup, irrespective of variations in the diode chip parameters, as depicted in Fig. \ref{fig4}a; (2) The laser linewidth reaches 24 kHz, which is significantly narrower than the hundreds-of-kilohertz linewidths typical of conventional ECDLs based on interference filters or gratings. Furthermore, the tuning range of the 75-mL AFL has been extended from 2 pm in conventional AFLs to 9 pm, enabling it to scan across the transition lines of both $^{85}$Rb and $^{87}$Rb without mode-hopping. Moreover, the combination of a compact structural design and a multi-stage temperature control system significantly suppresses the long-term wavelength drift of the 75-mL AFL. Over 42 days of free-running operation, the wavelength drift remained below 1.5 pm with no mode-hopping observed, as illustrated in Fig. \ref{fig4}b. When compared alongside an established commercial laser (Toptica TA pro) in the identical environment, the performance advantage is clear: even if the substantial wavelength drift of the TA pro during the initial 2–4 days is disregarded as a warm-up period, its wavelength drift during subsequent operation remains three times greater than that of the 75-mL AFL, as shown in Fig. \ref{fig4}c.

The principal contribution of the 75-mL AFL is its improved power stability compared with conventional AFLs. Operating the $\upmu$FADOF in the stable regime makes its transmittance insensitive to optical intensity variations induced by increasing the diode current. Meanwhile, as the diode current increases, the laser mode shifts in frequency. When it moves outside the transmission window, the laser switches to another longitudinal mode within the transmission window, producing the periodic behavior shown in Fig. \ref{fig4}a. Within each cycle, the frequency drift causes a slight decrease in $\upmu$FADOF transmittance, as indicated by the black curve in Fig. \ref{fig3}g. This decrease offsets the gain enhancement associated with the increasing diode current, resulting in the step-like power–current characteristic shown by the blue curve in Fig. \ref{fig4}d. A detailed analysis is provided in the Supplementary Material. Within each plateau of this profile, fluctuations in the final output power arising from diode current noise are suppressed. 

As a result, when benchmarked against traditional AFLs (orange line in Fig. \ref{fig4}d), the 75-mL AFL exhibits a two-order-of-magnitude enhancement in power stability, especially over long-term operation. Nevertheless, to fully realize the power stability potential of the 75-mL AFL, it is essential to investigate the effects of internal and external parameters—such as the FADOF temperature, laser mode drift, and ambient temperature—on its output power, and to optimize these parameters accordingly.

As illustrated in Fig. \ref{fig4}e, we investigated the influence of laser mode drift on the output power under various FADOF temperatures; in this plot, darker colors indicate smaller power variations induced by parameter fluctuations. Three optimal operating regions emerge, each offering distinct advantageous characteristics. Region 1 corresponds to a FADOF temperature range of 86–88°C, where FADOF temperature fluctuations exert the minimal impact on power when the laser output mode wavelength is at 780.242 nm. Region 2 corresponds to an output wavelength range of 780.241–780.242 nm, where output mode drift has the least effect on power when the FADOF temperature is fixed at 86°C. Region 3 spans an output wavelength range of 780.2432–780.2445 nm and a FADOF temperature range of 84–85°C; within this zone, both FADOF temperature fluctuations and output mode drifts have a relatively minor impact on power. Considering that the $^{85}$Rb transition lines fall within the 780.2437–780.2441 nm range, we configured the 75-mL AFL to operate in Region 3 to satisfy the subsequent requirements for a quantum device laser source.

Following a similar principle, we conducted a comprehensive optimization of the diode current (C), FADOF temperature (T), laser mode frequency (F), and ambient temperature (T). The resulting laser power instability, as plotted in Fig. \ref{fig4}f, reached $\mathbf{9 \times 10^{-6}}$ at 1 s and $\mathbf{3.2 \times 10^{-5}}$ at 8400 s. Compared to conventional AFLs, this represents a two-order-of-magnitude improvement in short-term power instability and a three-order-of-magnitude improvement in long-term performance. During a 12-hour free-running test, the output power fluctuation remained below 100 nW, yielding a normalized root-mean-square (RMS) stability of 0.0084\% and a peak-to-peak stability of 0.028\% (Fig. \ref{fig4}g). This resolves the longstanding issue of unstable output power in AFLs.

\vspace{3pt}
\noindent
\textbf{Application demonstrations} \\
\noindent 
We implemented an MTS-locked compact Faraday optical frequency standard (CFOFS) based on the 75-mL AFL, as illustrated in Fig. \ref{fig1}d. The CFOFS beats against a single-pass frequency-doubled optical comb locked to a hydrogen maser, enabling performance characterization and evaluation. The 75-mL AFL endows the CFOFS with turn-key functionality, resolving a universal challenge for optical atomic clocks: long-term local oscillator laser drift can cause unlocking or mislocking, requiring manual intervention for recovery. In the face of severe fluctuations in diode parameters (such as current and temperature), the CFOFS—like previous AFL-based optical frequency standards—can automatically relock once the diode parameters stabilize, whereas optical frequency standards based on other lasers might fail to operate. 

Furthermore, under environmental temperature shocks, as shown in Fig. \ref{fig5}a (where the ambient temperature rises from 21°C to 30°C within 6 minutes), although the CFOFS output frequency experiences a drift on the order of 10 kHz during the initial phase of the shock, it remains firmly locked to the atomic transition. 7000 s after the onset of the temperature shock, the CFOFS frequency output regains stability, with a net frequency drift of merely 32 kHz across the entire thermal event. 

Regarding mechanical shocks, as depicted in the inset of Fig. \ref{fig5}b, vibration shocks (0.8–2.5 g) were applied every 60 s, inducing CFOFS unlocking but followed by automatic relocking within 3 s. Over 20 vibration cycles (Fig. \ref{fig5}b), the total laser frequency drift was 378 Hz, which is comparable to that of the CFOFS during 20 minutes of normal operation. This indicates that mechanical shocks of this magnitude barely affect the frequency performance of the 75-mL AFL-based MTS optical frequency standard.

Another benefit of the 75-mL AFL is the significant reduction in the long-term power instability of the local oscillator. Compared to the Voigt laser (VL) local oscillator used for MTS schemes \cite{liu2025turn}, the 75-mL AFL reduces long-term power instability by over two orders of magnitude, reaching $3.2 \times 10^{-5}$ at 8400 s. This overcomes the dominant limitation that power instability previously imposed on the frequency instability of compact optical frequency standards, achieving an instability metric of $3 \times 10^{-13}$ at 10000 s. As illustrated in Fig. \ref{fig5}c, the in-loop noise (dictated by the internal noise of the phase-locked loop) and the overall system temperature drift (driven by the overall CFOFS system temperature) have replaced power drift as the primary limiting factors for CFOFS frequency instability. Compared to other MTS-locked compact optical frequency standards, the 10000-s instability of the CFOFS achieves state-of-the-art levels. Specifically, when compared to compact optical frequency standards that do not utilize an AFL as their local oscillator \cite{lee2023laser}, the CFOFS demonstrates an improvement in long-term frequency stability by a factor of 3 to 4, as shown in Fig. \ref{fig5}d. Regarding short-term instability, the CFOFS requires further enhancement—such as mitigating residual amplitude noise in the MTS optical path and optimizing the frequency-locking circuitry—to ultimately reach the $10^{-14}$ level.

\vspace{6pt}
\noindent
\large\textbf{Conclusion} \\
\normalsize
\noindent 
In this study, we broke through the physical limits of AFL miniaturization and power stability, reducing its volume by a factor of 30 to a mere 75 mL, and improving its power instability by two orders of magnitude to $9 \times 10^{-6}$ at 1 s and $3.2 \times 10^{-5}$ at 8400 s. By investigating the atomic filtering phenomenon under the hyperfine Paschen-Back effect, we proposed a novel theoretical paradigm for atomic filters capable of generating an ultrabroadband, flat-top transmission window. Furthermore, by exploring the optical self-feedback mechanism of the atomic filter, we classified its optical operating regimes (based on optical intensity) into three distinct states: the saturable absorption regime, the stable regime, and the reverse saturable absorption regime. Operating in the stable regime vastly enhances the power stability of atomic filter devices, while the contrasting nonlinear transmission responses of the SA and RSA regimes suggest opportunities for intensity discrimination, optical limiting, and all-optical switching. Finally, we deployed the 75-mL AFL as the local oscillator for a quantum precision measurement device—a 780 nm compact MTS optical frequency standard. The AFL's inherent alignment to atomic transitions and its compact structural design endow the optical frequency standard with the capability to be field-deployed in complex environments. Meanwhile, its superior power and frequency performance substantially overcome the long-term frequency stability limitations of the optical frequency standard, improving this metric by a factor of 3 to 4 to $3 \times 10^{-13}$ at 10000 s. These results demonstrate the significance of AFLs in the design and development of field-deployable, high-performance quantum devices.

\vspace{6pt}
\noindent \textbf{Methods}\\
\begin{footnotesize}
\noindent 
\textbf{MTS} 

The output of the 75-mL AFL is split into two paths (see Fig. \ref{fig1}d). One path enters the modulation-transfer spectroscopy (MTS) optical setup, which utilizes a natural rubidium double-layer vapor cell as a reference, generating an MTS signal for the AFL frequency servo feedback. The second path beats against a single-pass frequency-doubled optical comb locked to a hydrogen maser, enabling comprehensive performance characterization and evaluation.

The locking procedure proceeds as follows: the 780 nm laser is split by a half-wave plate and a polarizing beam splitter (PBS) into pump and probe beams. The pump beam is phase-modulated by an electro-optic modulator (EOM), expanded to a 3-mm diameter by a collimating lens assembly, and then directed into the reference cell. It counter-propagates collinearly with the 1-mm-diameter probe beam, transferring the modulation signal onto the probe beam. This modulated probe beam is then captured by a photodetector and demodulated to generate an error signal. Finally, integrated control electronics automatically apply feedback to the diode current and the piezoelectric transducer (PZT) tuning voltage, locking the laser to the $^{85}$Rb D$_2$ transition line ($5^2S_{1/2}, F = 3 \rightarrow 5^2P_{3/2}, F' = 4$).

\vspace{3pt}
\noindent 
\textbf{75-mL AFL} 

The main components of the 75-mL AFL are a front‑facet AR‑coated laser diode, a collimating lens, a $\upmu$FADOF with a volume of 22.4 mL, and a cavity mirror with 80\% reflectivity mounted on a PZT. All elements are mounted together inside an invar base and temperature‑controlled from below by a TEC. Light emitted from the diode is collimated and sent through the $\upmu$FADOF for frequency selection, then reflected by the cavity mirror back along the same path to establish oscillation and produce the AFL output. To maximize mechanical stability, all adjustable mounts were eliminated and every component was fixed using threaded fittings and epoxy, yielding a minimalist structure without redundant adjustment.


\vspace{3pt}
\noindent 
\textbf{Theoretical calculations}

For a configuration with crossed polarizer and analyzer, the transmission is given, according to the theory of dispersive magneto-optical filters \cite{yeh1982dispersive}, by
\begin{equation}
T=\frac{1}{2}\exp\left(-\bar{\alpha}L\right)
  \left[\cosh\left(\Delta\alpha L\right)-\cos\left(2\rho L\right)\right]
\label{eq:transmission}
\end{equation}
Equation~\eqref{eq:transmission} can be rigorously decomposed into the product of an absorption term and a dispersion term:
\begin{equation}
T=T_{a}\times T_{d}
\label{eq:factorization}
\end{equation}
\begin{equation}
T_{a}=\frac{1}{2}\left[\exp\left(-\alpha_{+}L\right)+\exp\left(-\alpha_{-}L\right)\right]
     =\exp\left(-\bar{\alpha}L\right)\cosh\left(\Delta\alpha L\right)
\label{eq:Ta}
\end{equation}
\begin{equation}
T_{d}=\frac{1}{2}-\frac{\cos\left(2\rho L\right)}{2\cosh\left(\Delta\alpha L\right)}
\label{eq:Td}
\end{equation}
Here, $L$ is the length of the vapor cell, and $\bar{\alpha}$, $\Delta\alpha$ and $\rho$ are the average absorption coefficient, the circular dichroism coefficient and the optical rotation coefficient, respectively. These three quantities are related to the left- and right-circular susceptibilities by
\begin{equation}
\begin{aligned}
\bar{\alpha} &= \frac{\omega}{2c}\operatorname{Im}\left(\chi_{+}+\chi_{-}\right),\\
\Delta\alpha &= \frac{\omega}{2c}\operatorname{Im}\left(\chi_{+}-\chi_{-}\right),\\
\rho &= \frac{\omega}{4c}\operatorname{Re}\left(\chi_{+}-\chi_{-}\right)
\end{aligned}
\label{eq:definitions}
\end{equation}
It should be noted that, in practice, the atoms also exhibit Doppler broadening and a multilevel structure; the complete theoretical model and the corresponding derivations for the resulting transmission spectrum are presented in the companion paper \cite{xiao2026statedependentdiffusionspectrastrongly}. In the following, a simplified two-level model is adopted to illustrate the main scales of the intensity response.

For ease of analysis, we consider a two-level system, for which the left- and right-circular susceptibilities can be expressed as
\begin{equation}
\chi_{\pm}\propto\frac{\omega\mp\omega_{0}-i\gamma}
  {\left(\omega\mp\omega_{0}\right)^{2}+\gamma^{2}\left(1+s\right)}
\label{eq:susceptibility}
\end{equation}
where $\omega$ is the laser frequency relative to the central reference position, $\omega_{0}$ is the Zeeman shift of the left- and right-circular transitions, $\gamma$ is the Lorentzian linewidth of the atomic transition, and $s=I/I_{\mathrm{sat}}$ is the saturation parameter. Defining $r=\omega_{0}/\gamma$ (the ratio of the Zeeman shift to the natural linewidth), we consider the following two cases.

When the laser is at the line center, i.e., $\omega\approx 0$, the left- and right-circular susceptibilities are
\begin{equation}
\chi_{\pm}\propto\frac{\mp r-i}{\gamma}\frac{1}{1+s+r^{2}}
\label{eq:susceptibility-center}
\end{equation}
The corresponding coefficients are given by
\begin{equation}
\bar{\alpha}\propto-\frac{2}{\gamma}\frac{1}{1+s+r^{2}},\quad
\Delta\alpha=0,\quad
\rho\propto-\frac{2r}{\gamma}\frac{1}{1+s+r^{2}}
\label{eq:coefficients-center}
\end{equation}
Since $r^{2}\gg 1$ appears in the denominators of both $\bar{\alpha}$ and $\rho$, the absorption at the line center is weak at zero light intensity ($s=0$), and $T_{a}$ approaches unity. To ensure the highest transmission at the line center, the dispersion is tuned to the rotation angle $\rho L=\pi/2$, so that $T_{d}=1$. For $s\ll r^{2}$, a change in the light intensity hardly alters the average absorption coefficient or the optical rotation coefficient, and both $T_{a}$ and $T_{d}$ remain essentially unchanged. Only when $s$ becomes comparable to $r^{2}$---that is, when the saturation broadening becomes comparable to the Zeeman shift (the threshold of the reverse saturable absorption regime)---do $\bar{\alpha}$ and $\rho$ change appreciably: the reduced absorption brings $T_{a}$ closer to unity, while the dispersion deviates from $\pi/2$, causing $T_{d}$ to decrease significantly.

When the laser is in resonance with one side of the transition, i.e., $\omega\approx\omega_{0}$, the left- and right-circular susceptibilities are
\begin{equation}
\begin{aligned}
\chi_{+}
&\propto -\frac{i}{\gamma}\,
\frac{1}{1+s}, \\[4pt]
\chi_{-}
&\propto \frac{2r-i}{\gamma}\,
\frac{1}{1+s+4r^{2}}
\end{aligned}
\tag{10}
\label{eq:single_side_resonance}
\end{equation}
Accordingly, the coefficients become
\begin{equation}
\begin{aligned}
\bar{\alpha} &\propto -\frac{1}{\gamma}\left(\frac{1}{1+s}+\frac{1}{1+s+4r^{2}}\right),\\
\Delta\alpha &= -\frac{1}{\gamma}\left(\frac{1}{1+s}-\frac{1}{1+s+4r^{2}}\right),\\
\rho &\propto -\frac{2r}{\gamma}\frac{1}{1+s+4r^{2}}
\end{aligned}
\label{eq:coefficients-side}
\end{equation}
For $r^{2}\gg 1$ and $0<s<1$, the first term of $\bar{\alpha}$ is much larger than the second term, i.e., $\bar{\alpha}\approx\Delta\alpha\propto-\frac{1}{\gamma}\frac{1}{1+s}$; under these conditions the atomic absorption is relatively strong, and increasing $s$ causes the absorption to decay rapidly. At the same time, owing to the presence of $r^{2}$, increasing $s$ has almost no effect on $\rho$. Only when $s$ becomes comparable to $r^{2}$ does $T_{d}$ decrease significantly.

It is worth noting that, at low light intensities ($s\ll r^{2}$), the atomic absorption at the one-sided resonance is relatively strong ($T_{a}<1$) and weakens markedly as the intensity increases, with $T_{a}$ gradually approaching unity; at the line center, by contrast, the atomic absorption is intrinsically weak ($T_{a}$ close to $1$) and remains almost unchanged as the intensity increases. The overall absorption shown in Fig.~3c, which weakens everywhere as the intensity increases, is due to the large magnetic-field splitting, which makes resonant transitions present almost everywhere.

For the atomic dispersion, on the other hand, whether at the line center or at the one-sided resonance, it decreases significantly only when $s$ becomes comparable to $r^{2}$.

Accordingly, the evolution with light intensity can be roughly divided into three stages:
\begin{itemize}
\item[(i)] Saturable absorption (SA) regime: $s$ is comparable to $1$, i.e., the saturation broadening is comparable to the natural linewidth; the absorption term $T_{a}$ decreases significantly.
\item[(ii)] Reverse saturable absorption (RSA) regime: $s$ is comparable to $r^{2}$, i.e., the saturation broadening is comparable to the Zeeman shift; the dispersion term $T_{d}$ decreases significantly.
\item[(iii)] Stable regime: intermediate between the two, in which both the absorption term $T_{a}$ and the dispersion term $T_{d}$ vary only slowly.
\end{itemize}
Upper intensity limit of the saturable absorption regime: theoretically, it is defined by the intensity at which $T_{a}>99\%$ at the line center; experimentally, by the intensity at which the line-center transmission of the measured spectrum exceeds $82\%$ (owing to losses of the optical components). Lower intensity limit of the reverse saturable absorption regime: theoretically, it is defined by the intensity at which $T_{d}<99\%$; experimentally, by the intensity at which the line-center transmission falls below $82\%$. The stable regime lies between these two thresholds.

\vspace{3pt}
\noindent 
\textbf{Power instability optimization}


Targeted parameter optimization is essential to fully exploit the 75-mL AFL’s potential. Replacing traditional gratings or interference filters with an atomic filter as the frequency-selection element eliminates the sensitivity of laser frequency and power to mechanical angular variations. However, it introduces a new dependency: the laser power fluctuates with the FADOF temperature (as the FADOF transmission spectrum drifts with temperature). In the AFL design, improving active temperature control to restrict $\upmu$FADOF temperature fluctuations to below 2 mK, combined with operating the AFL within the optical “stable regime”, largely mitigates the noise induced by $\upmu$FADOF temperature drifts. Furthermore, we fine-tuned the $\upmu$FADOF temperature in 0.1°C increments to monitor variations in the AFL output power, selecting an operating point where the rate of power change is minimized (indicated by the darker regions in Fig. \ref{fig4}e). This step is denoted as T-Optimized. Subsequently, active temperature control was applied to the entire AFL package to isolate it from ambient temperature fluctuations. This improved the Allan deviation of the $\upmu$FADOF temperature instability to 0.19 mK at 1 s and 0.47 mK at 1000 s. Combined with the previous measure, this dual-temperature optimization is designated as TT-Optimized.

When the laser frequency drifts, any unevenness in the FADOF transmission profile causes fluctuations in transmittance, subsequently leading to unstable laser power. The AFL specifically employs an ultrabroadband, flat-top $\upmu$FADOF transmission profile, which drastically suppresses the power noise originating from such spectral unevenness. Moreover, the intrinsic gain profile of the laser diode chip also influences the output power. By fine-tuning the diode current in 0.1 mA increments and monitoring the resulting power variations, we selected an operating current that minimizes the rate of power fluctuation. For the AFL used in this experiment, the optimal driving current was determined to be 117.99 mA; this step is designated as C-Optimized. Next, by utilizing MTS to lock the laser to the target operating frequency, a frequency instability of $3 \times 10^{-13}$ at 1 s was achieved. This fundamentally eliminates power noise induced by laser frequency drift, with this step designated as F-Optimized.

In summary, beyond the AFL's inherent principle-level advantages in power stabilization, we implemented a comprehensive parameter optimization encompassing the diode current, FADOF temperature, overall AFL ambient temperature, and output frequency locking. Denoted collectively as CTTF-Optimized, this approach ultimately achieved a power stability of $9\times10^{-6}$ at 1 s and $3.2\times10^{-5}$ at 8400 s, as depicted by the red line in Fig. \ref{fig4}f. For higher performance demands, an active power stabilization module can be integrated, enabling potential breakthroughs to $10^{-7}$ short-term stability and $10^{-8}$ long-term stability.

    
\vspace{6pt}
\noindent\textbf{Data availability}\\
The data supporting the plots and other findings of this study are available from the corresponding authors upon reasonable request.

\vspace{6pt}
\noindent\textbf{Code availability}\\
The code supporting the findings of this study is available from the corresponding authors upon reasonable request.

\end{footnotesize}
\vspace{20pt}

%
%

\clearpage

\setcounter{figure}{0}
\renewcommand{\thefigure}{S\arabic{figure}}

\vspace{6pt}
\noindent\large\textbf{Supplementary Material}\\
\normalsize
\noindent\textbf{Step-like power-current characteristics of the 75-mL AFL}\\
\noindent

As the laser-diode current increases, the lasing longitudinal mode redshifts. Once the mode drifts beyond the transmission window of the $\upmu$FADOF, the cavity loss increases and mode hopping is triggered, which switches the laser to the longitudinal mode with the largest net intracavity gain. Each mode hop therefore terminates the current operating step and initiates the next, giving rise to the step-like power-current (P-I) characteristics described in Fig.~S3. During a current scan, the maximum mode-hop-free tuning range is jointly determined by the diode gain spectrum and the $\upmu$FADOF transmission spectrum, and is generally smaller than the cavity free spectral range (FSR = 4.55 GHz). When the laser just starts to oscillate (48 - 80 mA), the $\upmu$FADOF gradually evolves from the saturated-absorption (SA) regime to the stable regime, accompanied by a pronounced variation in its transmission spectrum. The resulting positive optical feedback gives the first operating step (Step 1) the widest tunable range (780.2394-780.2465 nm), as shown in Fig.~S1. In the subsequent steps, the $\upmu$FADOF is already in the stable regime, so the tunable range remains nearly constant from step to step.

\begin{figure}[htbp]
\centering
\captionsetup{singlelinecheck=no, justification = RaggedRight}
\includegraphics[width=0.5\textwidth]{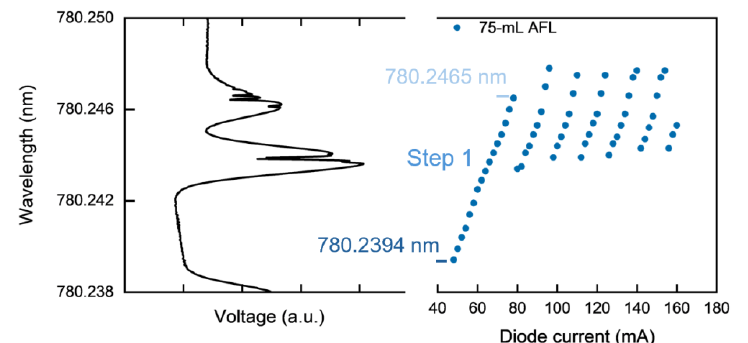}
\caption{Saturated-absorption spectrum of rubidium and the $\lambda$-I curve of the 75-mL AFL.}
\label{fig:S1}
\end{figure}

To explain the power evolution in Step 1 as the laser-diode current increases, two factors must be considered: the increase in diode gain and, more importantly, the change in $\upmu$FADOF transmission induced by the drift of the lasing mode. As shown in Fig.~S2, the $\upmu$FADOF operates mainly at intensities between 1000 and 5000 mW/cm$^2$ in Step 1. As the laser wavelength scans from 780.2394 to 780.2465 nm, the $\upmu$FADOF transmission first decreases with a very small, or nearly zero, slope, leaving the cavity loss almost unchanged; together with the increase in diode gain, this produces an approximately linear rise in laser output power over 780.2394-780.2437 nm. The transmission then decreases more steeply, so the intracavity loss increases and nearly cancels the diode gain increase; as a result, the net intracavity gain remains almost constant, and the P-I curve exhibits a plateau over 780.2437-780.2460 nm. Subsequently, within the 780.2460-780.2475 nm interval of the transmission spectrum, the $\upmu$FADOF transmission remains nearly constant, so the net gain rises again and the laser output power continues to increase. The 780.2475 nm boundary is identified from the transmission spectrum at 1000 mW/cm$^2$; at 5000 mW/cm$^2$, this boundary shifts toward shorter wavelengths (Fig.~S2). Once the wavelength exceeds this boundary, the $\upmu$FADOF transmission falls rapidly, and the output power decreases rapidly until mode hopping occurs.

\begin{figure}[tbp]
\centering
\captionsetup{singlelinecheck=no, justification = RaggedRight}
\includegraphics[width=0.4\textwidth]{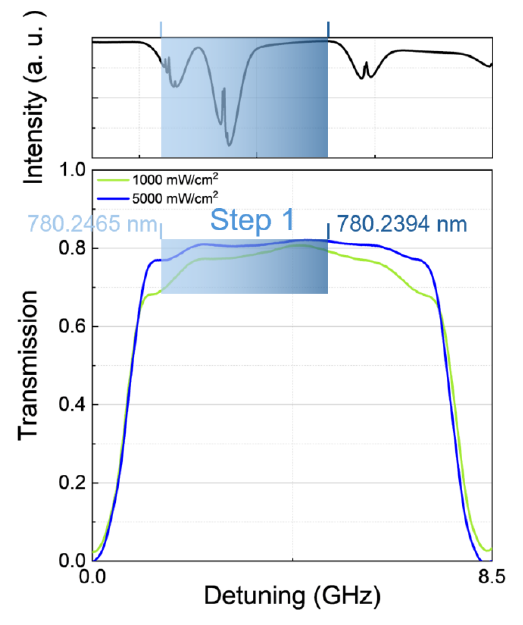}
\caption{Saturated-absorption spectrum of rubidium and the $\upmu$FADOF transmission spectrum with the 75-mL AFL in operation.}
\label{fig:S2}
\end{figure}

As shown in Fig.~S3, Step 1 undergoes mode hopping at 780.2465 nm and therefore displays only a rise (780.2394-780.2437 nm)-plateau (780.2437-780.2460 nm)-rise (780.2460-780.2465 nm) sequence. Steps 2-4 span the wavelength range 780.2434-780.2478 nm; their two boundaries are close to the lower-wavelength edge of the plateau (780.2437 nm) and the onset of the declining region (780.2475 nm), respectively, so these steps are dominated by a plateau (780.2434-780.2460 nm)-rise (780.2460-780.2478 nm) sequence. Steps 5 and 6 span the wavelength range 780.2440-780.2477 nm. At the higher optical intensity, the boundary of the final declining region shifts from 780.2475 nm to 780.2466 nm, so these steps display a plateau (780.2440-780.2460 nm)-rise (780.2460-780.2466 nm)-decline (780.2466-780.2477 nm) sequence.

Overall, the latter portion of Step 1 and the plateau regions of Steps 2 and 3 span wider diode-current intervals and are therefore the most suitable operating ranges for stable laser output.

\begin{figure}[tbp]
\centering
\captionsetup{singlelinecheck=no, justification = RaggedRight}
\includegraphics[width=0.5\textwidth]{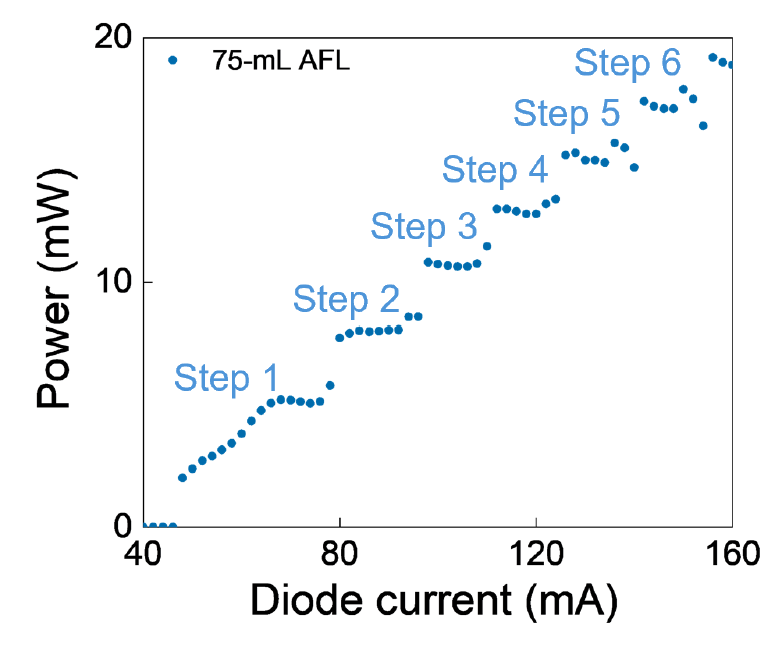}
\caption{P-I curve of the 75-mL AFL.}
\label{fig:S3}
\end{figure}

\bibliography{Ref.bib}


\vspace{12pt}
\begin{footnotesize}

\vspace{6pt}
\noindent \textbf{Acknowledgment}

\noindent This research was funded by the Quantum Science and Technology - National Science and Technology Major Project (2021ZD0303200), National Natural Science
Foundation of China (62405007), and Hebei Provincial Natural Science Foundation Basic Research Special Project - 2025 Basic Research Program Proof-of-Concept Project (F2025109009).
\noindent

\vspace{6pt}
\noindent \textbf{Author contributions}
  
\noindent  Z.L., T.S., A.D. and J.C. conceived the ideas and designed the experiments. T.S., A.D. and J.C. supervised the work. Z.L. and Z.X. performed the theoretical calculations. Z.L., Z.W., X.G., X.Q. and S.W. performed the experiments. Z.L. and Z.X. analysed the data. Z.L., Z.X., H.S., B.W., J.Z., X.X., T.S. and J.C. wrote the paper. All authors discussed the results and commented on the paper.

\vspace{6pt}
\noindent
\textbf{Additional information} 

\noindent Supplementary information is available in the online version of the paper. Reprints and permissions information is available online. Correspondence and requests for materials should be addressed to T.S. and A.D.

\vspace{6pt}
\noindent \textbf{Competing financial interests} 

\noindent The authors declare no competing financial interests.
\end{footnotesize}

\end{document}